\documentclass[aps,prx,twocolumn,final,letterpaper,floatfix,superscriptaddress]{revtex4-2}
\usepackage{appendix}
\usepackage{graphicx}   
\usepackage{import}
\usepackage{comment}
\usepackage{epstopdf}
\usepackage{bm}
\usepackage{amssymb}
\usepackage{quotes}
\usepackage{braket}
\usepackage{transparent}
\usepackage{dcolumn}
\usepackage{multirow}
\usepackage{mdframed}
\usepackage{booktabs}
\usepackage{color}
\usepackage{bm}
\usepackage{dsfont}
\usepackage{slashed}
\usepackage{enumitem}
\usepackage{amsmath} 
\usepackage{changepage} 
\usepackage{hyperref}
\usepackage{siunitx}
\usepackage{etoolbox}

\makeatletter

\def\@author@join@#1#2#3{%
  \@if@empty{#2}{%
    \def\@author{{#1}{\frontmatter@footnote{#3}}}%
  }{%
    \def\@author{{#1}{#2%
      \textsuperscript{,\,}\frontmatter@footnote{#3}}}%
  }%
}

\patchcmd{\doauthor}
  {\frontmatter@footnote{#2}}
  {#2}
  {}
  {\PackageError{authornotes}
    {Could not patch REVTeX author notes}
    {Check your REVTeX version.}}

\makeatother
\hypersetup{
    colorlinks=true,
    citecolor=blue, 
    linkcolor=blue, 
    urlcolor=blue   
}

\begin{document}
\rmfamily

\title{Enhanced light extraction in roughened plastic scintillators}

\author{William Michaels}
\thanks{These authors contributed equally to this work.}
\email{willmich@mit.edu}
\affiliation{Research Laboratory of Electronics, MIT,
  Cambridge, MA 02139, USA}
\affiliation{Department of Electrical Engineering and
  Computer Science, MIT, Cambridge, MA 02139, USA}

\author{Amelie Chan}
\thanks{These authors contributed equally to this work.}
\affiliation{Research Laboratory of Electronics, MIT,
  Cambridge, MA 02139, USA}
\affiliation{Department of Materials Science and Engineering,
  MIT, Cambridge, MA 02139, USA}

\author{Sachin Vaidya}
\email{svaidya1@mit.edu}
\affiliation{Research Laboratory of Electronics, MIT, Cambridge, MA 02139, USA}
\affiliation{Department of Physics, MIT, Cambridge, MA 02139, USA}

\author{Simo Pajovic}
\email{pajovics@caltech.edu}
\affiliation{Department of Mechanical Engineering, MIT, Cambridge, MA 02139, USA}
\affiliation{Department of Applied Physics and Materials Science, California Institute of Technology, CA 91125, USA}

\author{Joshua~Chen}
\affiliation{Research Laboratory of Electronics, MIT, Cambridge, MA 02139, USA}
\affiliation{Department of Electrical Engineering and Computer Science, MIT, Cambridge, MA 02139, USA}

\author{Charles Roques-Carmes}
\affiliation{Research Laboratory of Electronics, MIT, Cambridge, MA 02139, USA}
\affiliation{Edward L. Ginzton Laboratory, Stanford, CA 94305, USA}
\affiliation{Institute of Science and Technology Austria, Klosterneuburg, Austria}

\author{Steven E. Kooi}
\affiliation{Institute for Soldier Nanotechnologies, MIT, Cambridge, MA 02139, USA}

\author{Marin Solja\v{c}i\'{c}}
\email{soljacic@mit.edu}
\affiliation{Research Laboratory of Electronics, MIT, Cambridge, MA 02139, USA}
\affiliation{Department of Physics, MIT, Cambridge, MA 02139, USA}
\begin{abstract}
Scintillators are key components in X-ray imaging and particle detection, but their brightness in many applications is limited by total internal reflection, which prevents light from exiting the bulk where it is generated. Prior nanophotonic approaches to improving light extraction have relied on highly ordered structures fabricated using techniques that may be slow, expensive, and limited to small surface areas. Here, we show that mechanically roughening plastic scintillators to introduce random surface features enhances light outcoupling by 25\% while keeping the resolvable feature size below 0.2 mm. Our roughening approach is fast, inexpensive, and readily scales to large areas. We develop a parameter-free Monte Carlo optical scattering model based on microfacet theory that quantitatively predicts both the outcoupling enhancement and the imaging resolution across all roughness conditions. We further predict and observe remote outcoupling of emitted light, in which waveguided light exits the scintillator from a roughened region that is physically separated from the X-ray illumination. Our results indicate that surface roughness can enhance scintillator performance and that a straightforward model can explain a variety of experiments without relying on fitted parameters. 
\end{abstract}
\maketitle


\section{\label{sec:level1}Introduction}
Scintillators are widely used in applications such as medical imaging, geological surveying, nondestructive testing, and particle physics detection \cite{swank1954characteristics, lecoq2016development}. In each context, the purpose of the scintillator is to convert incident high-energy particles into visible light that can be more readily controlled and detected. A fundamental limiting factor in scintillator performance is the efficiency with which this generated light can be extracted from the material. Because scintillators are optically dense, photons emitted isotropically within the bulk are subjected to total internal reflection at the scintillator surface, reducing the signal for any downstream measurement. Recently, nanophotonics has been used to address this problem by shaping the light exiting scintillators at the wavelength scale, enabling control over the scattering amplitude and spectrum of emitted light, as well as diffraction \cite{roques2022framework, martin2025large, choi2026hybrid, choi2026supercollimating}. Common to these nanophotonic approaches is the use of highly ordered patterns that are commonly fabricated using nanofabrication techniques such as electron-beam lithography or focused ion beam milling \cite{knapitsch2011photonic, modrzynski2016light, liu2020large}. Such methods typically require extensive process development and optimization, which has so far restricted the widespread application of these demonstrations.

However, nanophotonic structures need not be highly ordered to produce scintillation enhancement. In other applications, random disorder can enhance certain photonic figures of merit. For instance, it has long been established that introducing surface disorder to solar cells can enhance light absorption \cite{yablonovitch1982statistical, yu2010fundamental}. In this case, surface disorder increases the likelihood that incoming photons will strike the surface within the acceptance cone, allowing them to make it into the active layer to be absorbed. Multiple studies have confirmed the efficacy of this approach and quantified its effect on device efficiency \cite{ma2012influences, kim2012effect, ctualu2017micromorphology, vafa2026balancing}. This mechanism is general and does not depend on material-specific properties, suggesting that the same mechanism should enhance outcoupling of emission via electromagnetic reciprocity. Indeed, surface roughness has been used to enhance light extraction in light-emitting diodes \cite{fujii2004increase, sung2019light, liu2026light}. Furthermore, introducing random surface disorder can be done quickly and cheaply, with the process readily scaling to large areas. In contrast to precise nanofabrication, mechanically abrasive techniques can process samples of arbitrary size in seconds with few requirements on material chemistry or process optimization.

\begin{figure*}
    \includegraphics[scale=0.75]{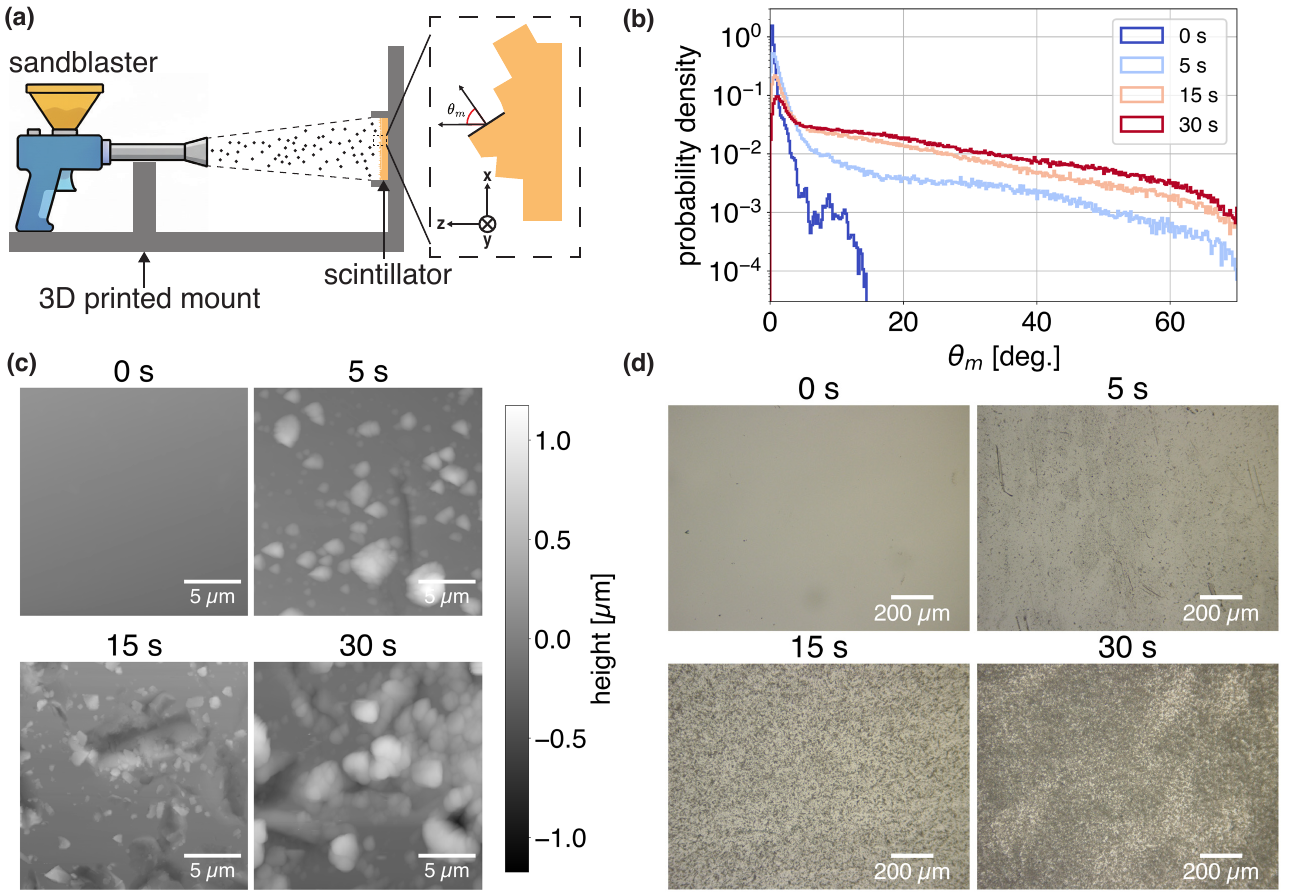}
    \caption{\textbf{Scintillator surface roughening via sandblasting.} \textbf{(a)} Experimental surface roughening setup. A 3D-printed mount holds the sandblaster at a fixed distance from the scintillator during roughening. For all roughened samples in this work, the distance between the tip of the sandblaster and the sample was 17.5 cm. The widths of the scintillator samples were approximately 2.5 cm. The inset shows the definition of the microfacet angle. \textbf{(b)} Microfacet angle probability density functions for different sandblasting times. $\theta_m$ is calculated using Eq. \ref{eq:1}. \textbf{(c)} Examples of AFM images used to compute the microfacet distributions. Note that multiple AFM images of each condition were aggregated to create each microfacet distribution. See Supplementary Note 2 for the full set of AFM images used in this work and their root mean square roughness values. \textbf{(d)} Optical microscope images of scintillator surface subjected to different roughening times.} 
    \label{fig:1}
\end{figure*}

The effect of surface roughness on the scintillator performance has been studied, mostly in the context of positron emission tomography (PET), where the relevant figures of merit are coincidence time resolution and depth-of-interaction resolution \cite{kilimchuk2009study, kilimchuk2010study, bircher2012investigation, berg2015optimizing, xie2019ultra}. These PET studies motivate examining how surface roughening affects both light collection and spatial resolution of scintillators used in X-ray imaging. In that context, a tradeoff exists: surface roughness may scatter photons as they exit the scintillator, blurring the image even as it enhances the total light yield due to broadening the escape cone. Modeling and characterizing this tradeoff is therefore essential to any practical application of surface roughening to X-ray imaging detectors.

In this work, we demonstrate the enhancement of light outcoupling from plastic scintillators using surface roughness. Through X-ray imaging experiments, we show that a simple sandblasting technique can create surface roughness that enhances light outcoupling by 25\% while maintaining imaging resolution below 0.2 mm. In addition, we predict and experimentally measure a phenomenon whereby scintillation light is remotely outcoupled by surface roughness physically separated from the incident high-energy illumination. To predict the performance of our roughened scintillator, we develop a straightforward model for light outcoupling without any free parameters that closely matches our experimental results. The same underlying model explains each of our experiments by changing only the locations of the illumination and surface roughness. This work shows how, with a relatively simple process, light extraction from scintillators can be predictably improved at scale.

\section{Sandblasting plastic scintillators}
We utilize a polyvinyltoluene-based plastic scintillator that emits red light at a peak wavelength of 580 nm \cite{SaintGobain2005BC430}. We chose a plastic scintillator because it is softer and easier to roughen compared to hard, inorganic crystal scintillators. Furthermore, plastic scintillators are used widely in applications such as particle detection, cargo inspection, and natural resource exploration \cite{beaulieu2016review, hajagos2018high}. To roughen them, we used a commercial sandblaster loaded with aluminum oxide grains (Paasche). The mechanical impact of these grains deforms and removes material on the surface of the scintillator, resulting in surface roughness. Figure \ref{fig:1}(a) shows a schematic of the experimental roughening setup used in this work. We used a custom 3D-printed mount to hold the sandblaster and sample at a fixed distance from one another. For more details on the roughening procedure, see Supplementary Note 1. 
\begin{figure*}
    \centering
    \includegraphics[scale=0.82]{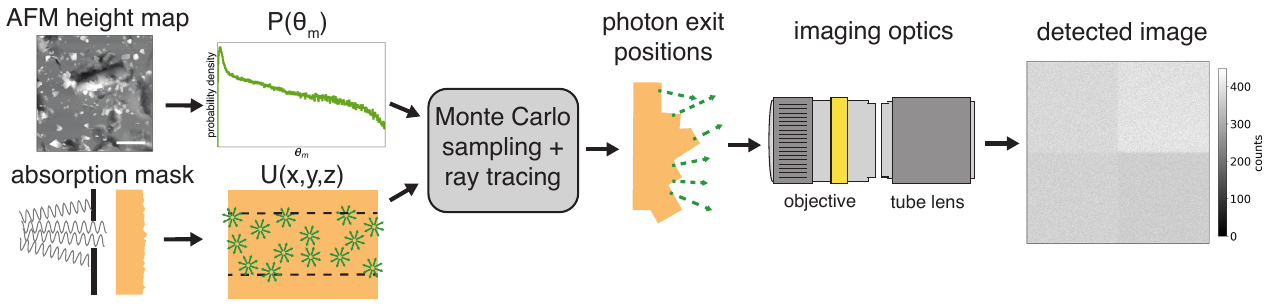}
    \caption{\textbf{Scintillator surface roughness simulation framework.} Schematic diagram of simulation pipeline used in this work. To simulate a given experiment, the AFM height map of the sample and the absorption mask for the experiment are used as inputs. These are converted to probability distributions for the microfacet angle and photon origin locations, respectively. A Monte Carlo model then generates many isotropic photon emission events and performs ray tracing to determine where photons will exit the scintillator. The final image on the detector is computed by simulating light collection by the imaging objective and applying any blurring from the optics as necessary.} 
    \label{fig:2}
\end{figure*}

Successful sandblasting is indicated by the increasing opacity of the sample with increased roughening time (without roughening, the sample is transparent). This can be seen with the naked eye; increasing disorder with roughening time is confirmed by eye as shown in Supplementary Note 1 and via optical microscopy as shown in Figure \ref{fig:1}(d). However, optical microscopy does not offer enough imaging resolution and topographic information to precisely quantify the surface roughness. Therefore, we used atomic force microscopy (AFM) to image and quantify the surface roughness of the samples as a function of roughening time. Since the field of view of AFM is small (20~\unit{\micro\meter} $\times$ 20~\unit{\micro\meter}), we collected multiple AFM images from different locations on the sample surface for each condition. Representative AFM images for each condition are shown in Figure \ref{fig:1}(c). We aggregated these images and computed local slope distributions using the formula
\begin{equation}\label{eq:1}
    \theta_m = \arctan\left(\sqrt{\left(\frac{\partial z}{\partial x}\right)^2 +\left(\frac{\partial z}{\partial y}\right)^2}\right),
\end{equation}
where $x$ and $y$ are the lateral positions in the image and $z$ is the height registered in the topographic map. The local surface normal is proportional to \((-\partial z/\partial x,-\partial z/\partial y,1)\), so its angle relative to the nominal surface normal satisfies \(\tan\theta_m=|\nabla z|\), yielding Eq. \ref{eq:1}. This angle characterizes the local interface orientation that determines a photon’s incidence angle and hence its reflection or transmission probability. In Figure \ref{fig:1}(b) we plot the microfacet angle distributions for different sandblasting times. We observe that longer sandblasting times create more microfacets at extreme angles due to the buildup of sand grain impacts on the scintillator surface. See Supplementary Note 2 for the complete set of AFM images used in this work as well as their root mean square roughness values. 
\section{Monte Carlo model of surface roughness}
\begin{figure*}
    \centering
    \includegraphics[scale=0.6]{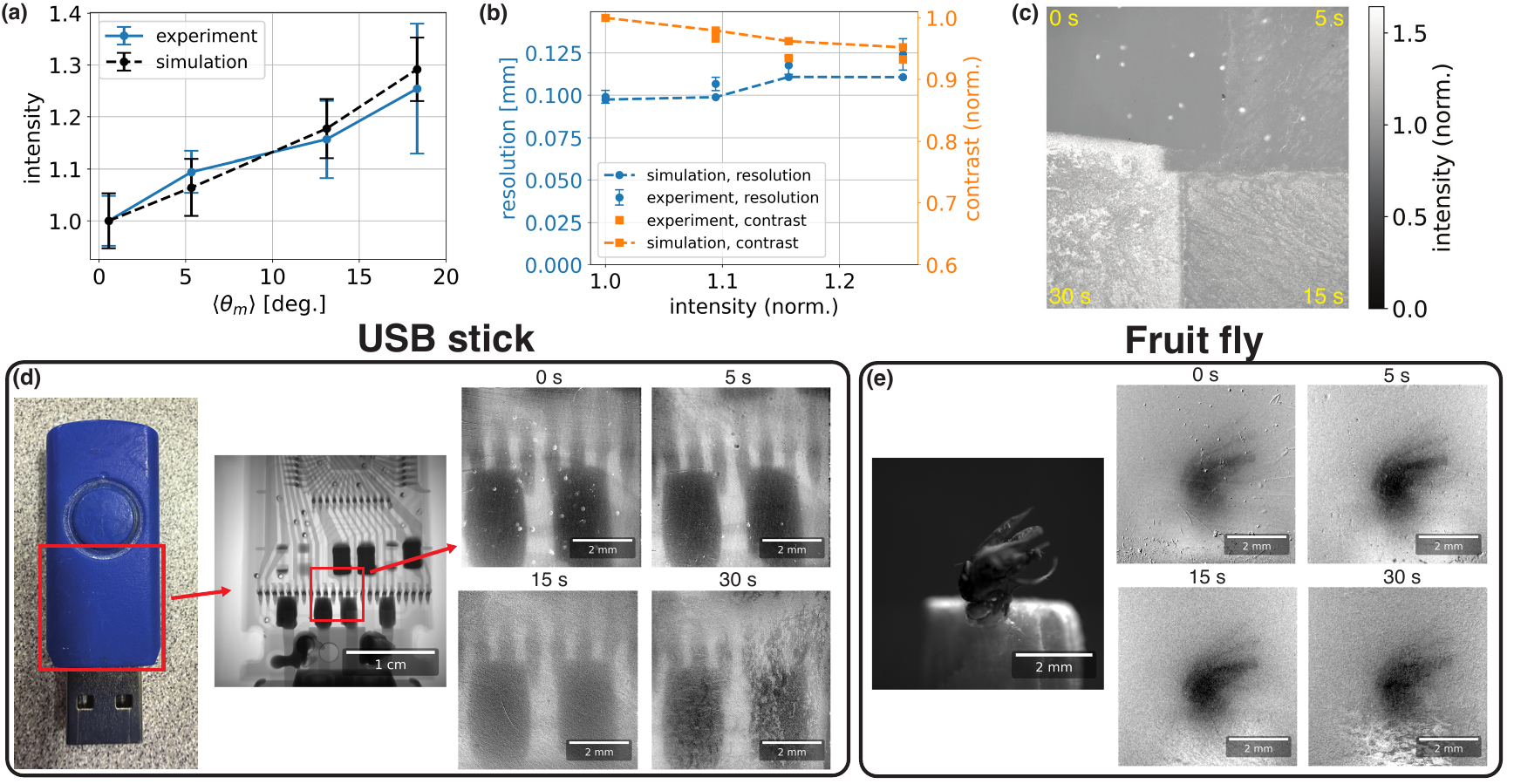}
    \caption{\textbf{Tailoring light extraction and spatial resolution of scintillators using surface roughness.} \textbf{(a)} Experimental (blue) and simulated (black) outcoupled intensity enhancement relative to an unroughened surface as a function of surface roughness. The error bars describe the spread in the intensity across the imaged region. $\langle\theta_m\rangle$ refers to the mean microfacet angle computed from AFM images of each surface (see Fig.~\ref{fig:1}). \textbf{(b)} Imaging resolution and image contrast as a function of the outcoupled intensity enhancement. For both the experiment and simulation, the resolution value is computed from fitting an error function to the edge spread function created by a stainless steel block in front of the scintillator. Contrast is computed from the same edge spread images by comparing the intensity in the dark and light regions. Contrast is normalized to the value for the unroughened scintillator. \textbf{(c)} Experimental X-ray image of scintillator surface. The time label in each region indicates the sandblasting time. The intensity is normalized so that the mean of the 0 s region is 1, as is done in (a). \textbf{(d)} Optical and X-ray images of USB stick internals. The leftmost image shows the outside of the USB stick taken with a smartphone camera. The center image was taken with a built-in inorganic scintillator in the micro-CT machine. On the right are X-ray images of specific metal components taken through our plastic scintillators with each of the roughness profiles. Images are flat-field corrected. \textbf{(e)} Optical and X-ray images of a fruit fly. On the left is an optical microscope image of the fruit fly taken in the micro-CT machine. The right panels show the same field of view captured through the roughened plastic scintillators. The experimental parameters for all data in this figure can be found in Supplementary Note 4.} 
    \label{fig:3}
\end{figure*}

While Monte Carlo models have been developed for scintillator surface roughness, they do not explore its effect on imaging performance nor provide an interpretable framework to explain how surface roughness affects overarching figures of merit \cite{janecek2010simulating, roncali2013simulation}. To this end, we developed a Monte Carlo model that accounts for emission, propagation, scattering, and subsequent collection of scintillated photons. An overview of the model is illustrated in Figure \ref{fig:2}. For a more detailed explanation of the model, see Supplementary Note 3. 

First, the scintillation process is modeled by the creation of a uniform distribution of dipoles that emit isotropically within the bulk. The dipole distribution is uniform along the propagation direction because the X-ray attenuation length of our plastic scintillator at energies of $\sim$ 10-100 keV is cm-scale, whereas the thickness of our samples is 0.5 mm. In other words, we assume no dependence of the X-ray absorption along the optical axis. The emitted photons are treated as randomly directed rays and propagated through the scintillator until they reach a surface. 

The interaction of the photons with each surface is modeled using microfacet theory \cite{walter2007microfacet}.  When a photon meets an interface, the angle it makes with the surface is computed using the corresponding local surface angle distribution derived from the appropriate AFM image. More precisely, the orientation of a given microfacet is defined by a polar angle and an azimuthal angle. We choose the polar angle from the measured microfacet angle distribution, while the azimuthal angle is chosen from a uniform distribution over [0, 2$\pi$). The microfacet angle distribution is also assumed to be spatially uncorrelated across the sample. To account for the fact that the photon's incident angle affects the probability of it striking microfacets with different angles, a shadowing-masking function is included when sampling from each distribution \cite{heitz2014understanding}. 

The photon's direction after the surface interaction is modeled using the refractive index of the scintillator (1.58) and Snell's law. At each interaction with an interface the photon also has a chance of being reflected according to the appropriate Fresnel reflection coefficient. To simulate the experimental measurements taken, photons that escape the scintillator are accepted or rejected based on the angle with which they exit the surface. Only photons that exit within the acceptance angle set by the numerical aperture of the imaging objective are counted in the image. All physical dimensions and roughness distributions in the simulation are taken directly from experimental data. Therefore, our model needs no free parameters to explain the collected data. To ensure sufficient statistics, the number of photons for each simulation is increased until the result converges.

To use this model to predict different experiments, we simply change the locations of the X-ray illumination and the surface roughness according to the situation, using measured AFM data to describe the roughness. As is demonstrated in subsequent sections, our model explains the measured data accurately, suggesting that changes in optical properties due to surface roughness result principally from changes in the microfacet slope distribution on the surface of the sample. 

\section{Enhanced light outcoupling and imaging resolution}
To determine the optical effects of surface roughness on plastic scintillators, we measured the enhancement in outcoupling due to surface roughness and the effect of the roughness on imaging resolution under X-ray excitation. The effect of surface roughness on light outcoupling is shown in Figure \ref{fig:3}(a). To obtain this data, we roughened different parts of a single scintillator sample for different amounts of time. We then took an X-ray image of the surface of the scintillator (Figure \ref{fig:3}(c)) and computed the mean intensity in each region. The X-ray illumination was generated by an X-ray tube in a Zeiss Xradia 620 Versa micro-CT and the image was captured using a Mitutoyo 5$\times$ NIR objective with a numerical aperture of 0.14. See Supplementary Note 4 for the full details on the experimental X-ray setup. The measured mean intensity is plotted against the mean microfacet angle obtained from AFM in Figure \ref{fig:3}(a). The simulated results were obtained using the Monte Carlo framework described above. In particular, to generate the curve in Figure \ref{fig:3}(a), AFM data from each of the quadrants was input to the model, with scintillation events occurring uniformly throughout the volume to reflect uniform X-ray illumination.  

In Figure \ref{fig:3}(a-c) the intensity values are normalized to the mean of the unroughened quadrant. This means that the region roughened for 30 seconds outcouples 25\% more light than the unroughened region, with lower roughening times interpolating between these cases. The range of intensities present in each quadrant also varies as a function of the roughening time. In Figure \ref{fig:3}(a) the error bars represent the width of the distribution of intensity values in that quadrant, not an experimental uncertainty. As the scintillator surface gets rougher, this width increases due to the wider range of angles a rough surface scatters light into. 

\begin{figure*}
    \centering
    \includegraphics[scale=0.5]{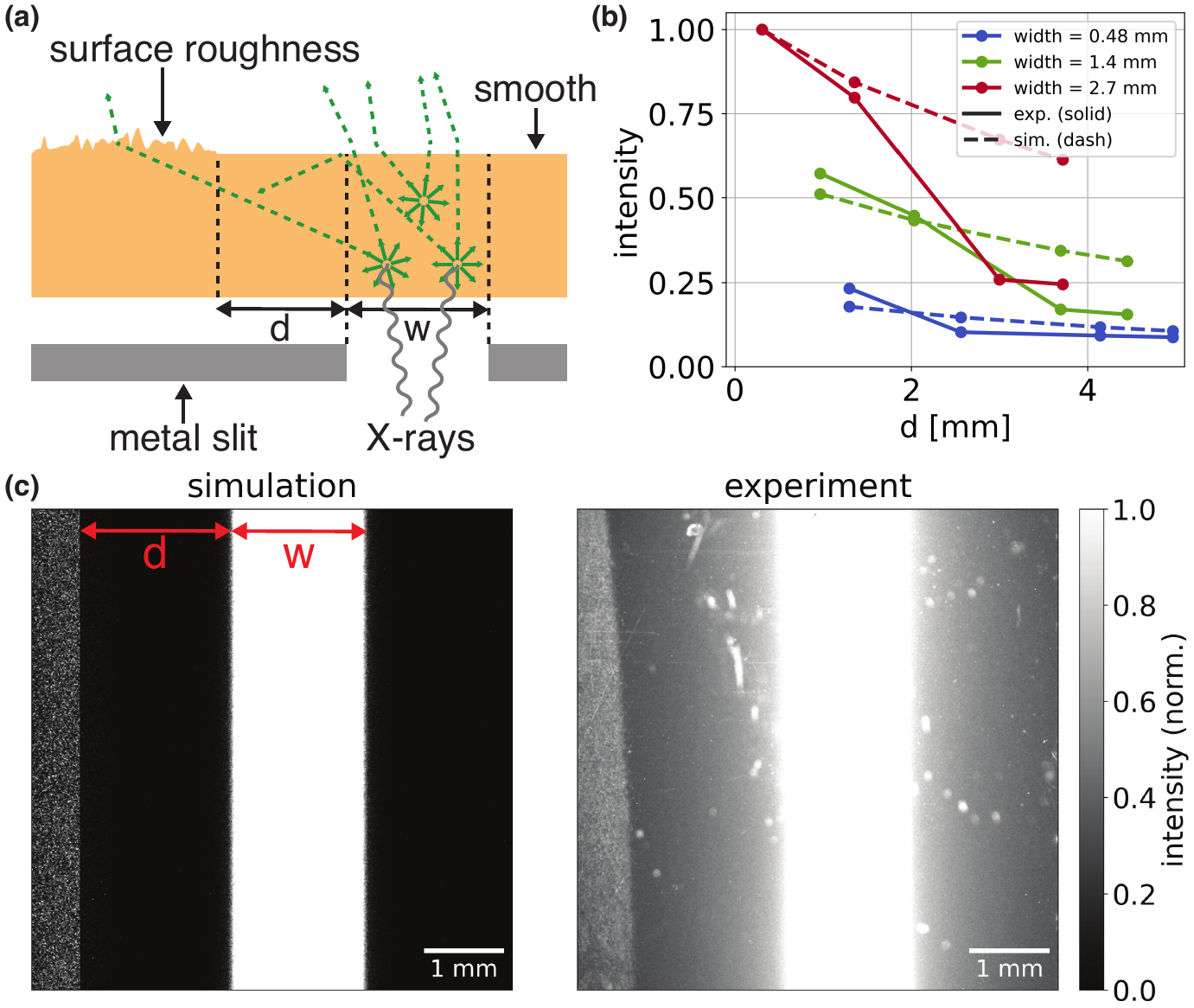}
    \caption{\textbf{Remote outcoupling from rough plastic scintillators.} \textbf{(a)} Schematic diagram of remote outcoupling experiment. $w$ refers to the width of the slit of illumination. $d$ refers to the distance of the slit from the rough region. \textbf{(b)} Experimental and simulated remotely outcoupled intensity as a function of slit width and distance from illumination. Intensity refers to the mean intensity in the rough region. \textbf{(c)} Sample comparison between the experimental and simulated remote outcoupling image. The condition shown is an illumination width of 1.4 mm and a distance from the illumination of 2 mm. Data are normalized to the maximum value in the entire dataset. All experimental data was captured with the X-ray tube voltage at 30 kV and the current at 67 $\mu$A.} 
    \label{fig:4}
\end{figure*}

In addition to outcoupling enhancement, one must consider the effect of surface disorder on the blurring of light to fully understand the implications of our surface roughening method on imaging. In particular, we expect some additional blurring from surface disorder due to the lateral spreading caused by additional scattering at the surface. To measure this effect, we imaged the edge of a stainless steel block through scintillators with different roughness profiles. By imaging a straight edge, we can fit an error function to the image profile and determine the amount of blurring occurring within the system. More precisely, based on the curve fit, we determine the corresponding line spread function, whose full width at half maximum we extract to approximate the imaging resolution. Comparing rough scintillators to smooth ones on this metric allows us to determine the effect of surface roughness. 

The results from this experiment are shown in Figure \ref{fig:3}(b). Here, we plot the imaging resolution against the relative outcoupling intensity for each sample from Figure \ref{fig:3}(a). We observe that while surface roughness degrades the imaging resolution of the scintillator, it does not do so severely. Even for the largest amount of roughness considered here, the minimum resolvable feature size remains below 0.2 mm, which is ample for common applications such as chest radiography \cite{huda2015x}. This demonstrates that it is possible to achieve notable outcoupling enhancement without compromising imaging capabilities through surface roughening. These trends hold for samples of the same material and another plastic scintillator subject to the same roughening conditions. See Supplementary Note 5 for data from other samples. To simulate the resolution experiment, photon generation was restricted to half of the scintillator, corresponding to the absorbing block placed in front of the scintillator during the experiment. The simulated image was processed in the same way as the experimental one to obtain the resolution values. 

Surface roughness also minimally degrades the image contrast. To understand this, we computed the absorption contrast in experiment and simulation using the images of the stainless steel block edge discussed above. Specifically, for each roughening condition we calculated the magnitude of the Weber contrast between the light and dark regions on either side of the edge \cite{peli1990contrast}. In Figure \ref{fig:3}(b) we plot the change in this quantity relative to the unroughened scintillator. We observe that there is a minimal reduction in the imaging contrast due to the surface roughness, indicating this is not a major drawback of introducing surface disorder. 

To confirm the minimal effect of roughness on imaging capabilities, we took X-ray images of internal USB stick structures and an iodine-stained fruit fly through our plastic scintillator samples. These images are shown in Figure \ref{fig:3}(d-e) along with optical and larger scale X-ray images. Notably, the small metal leads within the USB stick and the fruit fly wings remain visible for all roughness conditions after flat-field correction. Furthermore, image contrast is maintained as expected. Therefore, imaging performance is not significantly degraded when using a roughened scintillator. 

\section{Remote outcoupling}
Finally, we experimentally investigated a remote outcoupling effect predicted by our model. Remote outcoupling refers to outcoupling from a region of the scintillator physically separated from the X-ray illumination~\cite{pajovic2026volumetric}. This occurs due to waveguiding in the scintillator that directs light laterally through the sample. When this guided light reaches an area of surface roughness, it has an increased chance to outcouple towards the detector relative to the unroughened scintillator. This outcoupling is made more likely via the same mechanism of reducing total internal reflection as described in earlier sections. 

A schematic diagram of the setup we used to study this phenomenon is shown in Figure \ref{fig:4}(a). The schematic shows where X-ray illumination is restricted to and the separation $d$ between the illuminated region of width $w$ and the roughened region. In Figure \ref{fig:4}(b), we plot the mean intensity within the roughened region as a function of $d$ and $w$. The outcoupled intensity falls as the illumination region shrinks and the distance from the roughness to the illumination grows. These can be explained fairly simply. As the illuminated region shrinks ($w$ decreases), fewer X-ray photons are incident on the sample, and so fewer visible photons are generated through scintillation. As a result, the possible number of photons the roughened region can remotely outcouple also shrinks. As the distance to the roughness $d$ shrinks, visible photons are generated closer to the roughened region and therefore more likely to reach the rough region and be remotely outcoupled. The remote outcoupling measurements are also predicted by the Monte Carlo model, showing the ability of the model to describe a variety of situations. 

The simulated and experimental images for one of the cases in Figure \ref{fig:4}(b) are shown in Figure \ref{fig:4}(c). In both the simulated and experimental images, X-ray illumination is restricted to the bright white region in the center of the image. The entire surface of the scintillator is smooth, except for the small strip on the left side of each image. This region is roughened using the sandblaster for thirty seconds, the roughest condition considered in this work. As we can see in both the simulation and the experiment, light is outcoupled from the rough region despite it not being illuminated directly. This suggests that surface roughness can facilitate light outcoupling even if separated from the incident high-energy radiation.  

More generally, this phenomenon demonstrates that surface roughness can be used not only to increase the total extracted light, but also to control where that light preferentially leaves the scintillator. By placing roughened regions at selected locations, waveguided photons that would otherwise remain trapped can be extracted at designated collection sites, as has been demonstrated in light-emitting diodes \cite{park2007grating}. Such spatial control over outcoupling could enable new scintillator geometries and detector architectures in which optical collection is decoupled from the location of radiation absorption, providing additional flexibility in system design. The roughening procedure described in this work would make such schemes plausible since a simple mask such as a laser-cut piece of tape would be enough to generate any arbitrary shape with a roughened interior. Remote outcoupling has also recently been demonstrated with volumetric patterns in scintillators \cite{pajovic2026volumetric}. Our results show that the same principle can be used to control the location of outcoupling even if the pattern is constrained to the surface. 

\section{Discussion}
In this work we investigated the effect of surface roughness on the light yield, imaging resolution, and remote outcoupling ability of plastic scintillators. By utilizing a scalable sandblasting technique, we were able to increase the light outcoupling from our samples by 25\% without substantially degrading the achievable imaging resolution or contrast. We showed that this same surface roughness can lead to remote outcoupling, which enables light to be outcoupled from regions not illuminated directly by X-rays. To quantitatively explain our measurements, we developed a straightforward and interpretable model based on microfacet theory. 

This study highlights another application of surface roughness in photonics using a simple, cheap, and scalable roughening technique. While typical nanoscale patterning techniques require costly equipment and substantial time to tune the fabrication process, our roughening method utilizes inexpensive commercial tools and can be performed within minutes. In addition, our method is material-agnostic and scales to large areas, unlike most other nanophotonic interventions in scintillators.

While this work focuses on plastic scintillators, the principles underlying the outcoupling enhancement are material-independent, depending only on the geometry of the surface and the refractive index of the scintillator. This suggests that the same approach could be applied to inorganic scintillators such as LYSO, GAGG, or CsI:Tl, which are widely used in clinical PET and CT systems and which have refractive indices higher than the BC-430 used here. A higher refractive index implies a smaller critical angle for total internal reflection, meaning a larger fraction of emitted photons are trapped. This means there is a potentially larger absolute gain from surface roughening. To extend this work to these materials one must consider the different mechanical properties of inorganic scintillators, namely their increased hardness relative to plastic scintillators. It could also be fruitful to consider chemical or laser roughening techniques with these materials to achieve the same effect \cite{kurashige1998surface, cates2013increased}.

Beyond the effects demonstrated here, it remains an open question to what degree similar types of surface disorder could be used to manipulate other properties of the light exiting the scintillator. Previous studies of nanophotonic scintillators suggest that variables such as the spectrum, direction, and polarization of the outcoupled radiation can be controlled using nanophotonics \cite{roques2022framework}. Achieving such control with surface roughness would likely require more controlled roughness profiles. A simple computational model like the one developed here could likely aid in probing the necessary roughness properties to achieve a given effect.  

\section{Acknowledgments}
W.M. and J.C. acknowledge support from the National Science Foundation Graduate Research Fellowship Program under grant number 2637797. S.P. gratefully acknowledges support from the MathWorks Engineering Fellowship via MIT. C.R.C. is supported by a Stanford Science Fellowship. The authors acknowledge MIT SuperCloud and the Lincoln Laboratory Supercomputing Center for providing high-performance computing resources. This work was funded in part by the U. S. Army Research Office through the Institute for Soldier Nanotechnologies (ISN) at MIT, under Collaborative Agreement Number W911NF2320121. We also acknowledge support from the U.S. National Science Foundation under award number 2614673. We thank Christopher Hogan for help with operating and troubleshooting the experimental tools used in this work. 

\bibliography{main}

\clearpage
\onecolumngrid
\end{document}


\rmfamily

\title{Supplementary Information for: \\ Enhanced light extraction in roughened plastic scintillators}

\author{William Michaels}
\thanks{These authors contributed equally to this work.}
\email{willmich@mit.edu}
\affiliation{Research Laboratory of Electronics, MIT,
  Cambridge, MA 02139, USA}
\affiliation{Department of Electrical Engineering and
  Computer Science, MIT, Cambridge, MA 02139, USA}

\author{Amelie Chan}
\thanks{These authors contributed equally to this work.}
\affiliation{Research Laboratory of Electronics, MIT,
  Cambridge, MA 02139, USA}
\affiliation{Department of Materials Science and Engineering,
  MIT, Cambridge, MA 02139, USA}

\author{Sachin Vaidya}
\email{svaidya1@mit.edu}
\affiliation{Research Laboratory of Electronics, MIT, Cambridge, MA 02139, USA}
\affiliation{Department of Physics, MIT, Cambridge, MA 02139, USA}

\author{Simo Pajovic}
\email{pajovics@caltech.edu}
\affiliation{Department of Mechanical Engineering, MIT, Cambridge, MA 02139, USA}
\affiliation{Department of Applied Physics and Materials Science, California Institute of Technology, CA 91125, USA}

\author{Joshua~Chen}
\affiliation{Research Laboratory of Electronics, MIT, Cambridge, MA 02139, USA}
\affiliation{Department of Electrical Engineering and Computer Science, MIT, Cambridge, MA 02139, USA}

\author{Charles Roques-Carmes}
\affiliation{Research Laboratory of Electronics, MIT, Cambridge, MA 02139, USA}
\affiliation{Edward L. Ginzton Laboratory, Stanford, CA 94305, USA}
\affiliation{Institute of Science and Technology Austria, Klosterneuburg, Austria}

\author{Steven E. Kooi}
\affiliation{Institute for Soldier Nanotechnologies, MIT, Cambridge, MA 02139, USA}

\author{Marin Solja\v{c}i\'{c}}
\email{soljacic@mit.edu}
\affiliation{Research Laboratory of Electronics, MIT, Cambridge, MA 02139, USA}
\affiliation{Department of Physics, MIT, Cambridge, MA 02139, USA}

\noindent	

\noindent

\clearpage
\newpage

\renewcommand{\sp}{\sigma_+}
\newcommand{\sm}{\sigma_-}

\setlength{\parindent}{0em}
\setlength{\parskip}{.5em}
\vspace*{-2em}


\maketitle

\tableofcontents
\clearpage
\section{Scintillator roughening setup}
\begin{minipage}{\textwidth}
    \centering
    \includegraphics[width=0.9\textwidth]{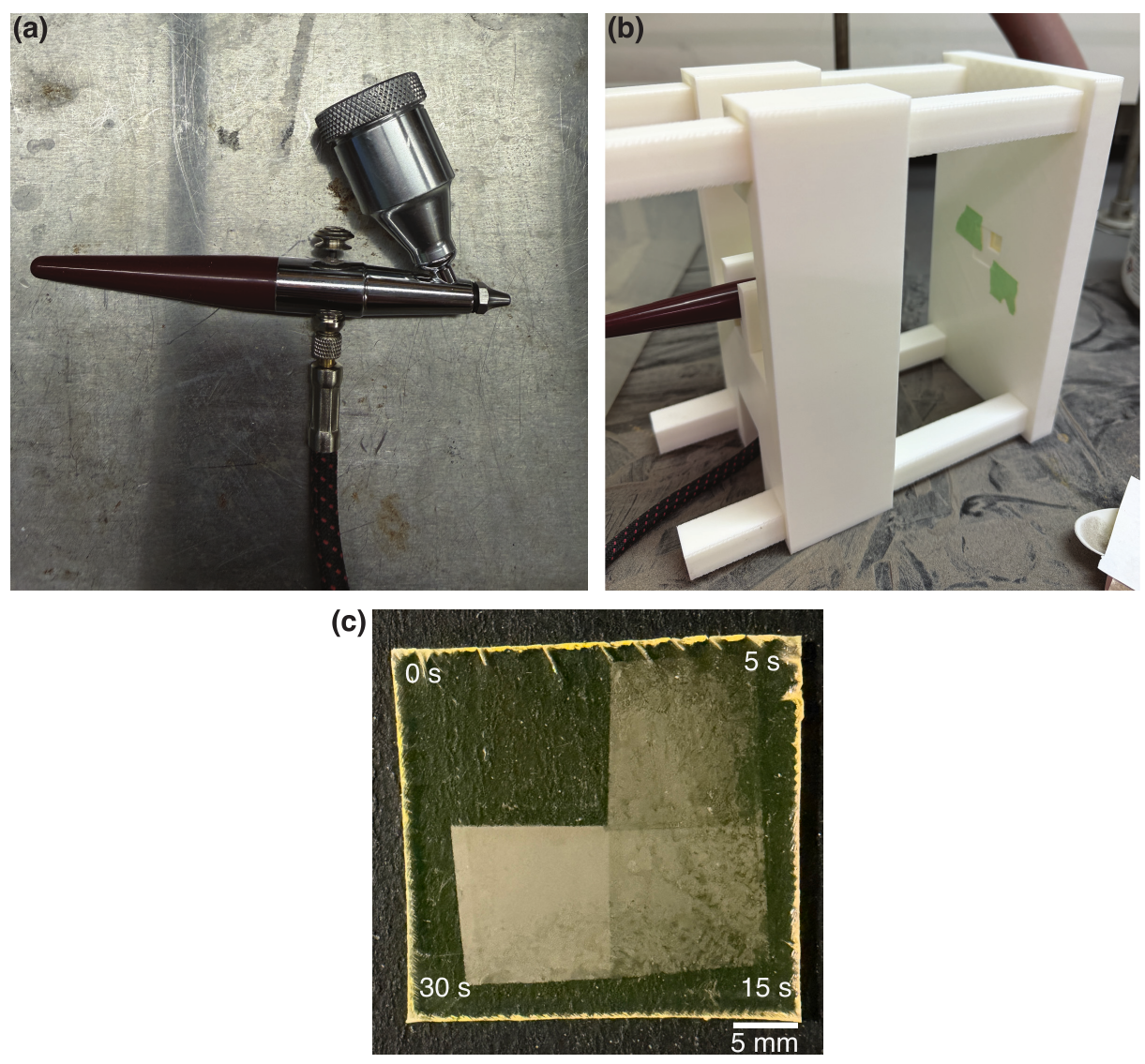}

    \captionof{figure}{
    \textbf{Scintillator roughening setup.}
    \textbf{(a)} Paasche sandblaster used during the roughening procedure. 
    \textbf{(b)} 3D-printed mount used to control variation between sandblasting trials. The sandblaster sits in the mount and can be moved towards and away from the scintillator sample that is secured in the mount using a 3D-printed cover. \textbf{(c)} Smartphone camera image of scintillator after roughening. Quadrant labels indicate the roughening times.}
    \label{fig:scintillator_roughening}
\end{minipage}
\vspace{1mm}

The material used in the main text is a polyvinyltoluene-based plastic scintillator that emits red light at a peak wavelength of 580 nm (BC-430, Luxium Solutions). BC-428 which emits at a peak wavelength of 480 nm was used in Supplementary Note 5 to confirm reproducibility. All scintillator samples were 0.5 mm thick and cut into 2.5 cm $\times$ 2.5 cm squares using scissors. To roughen the samples, we used a commercial sandblaster (Paasche) to bombard the samples with sand grains. The scintillator was held 17.5 cm from the tip of the sandblaster using a custom 3D-printed mount. This mount also ensured that the sand grain trajectories were perpendicular to the surface of the scintillator. See Figure \ref{fig:scintillator_roughening} for images of the sandblaster and the mount. 

We empirically observed that the flux of sand grains launched by the sandblaster changed as the hopper emptied. To ensure the flux of sand grains incident on the sample was constant, roughening was divided into five-second intervals, in between which the hopper of the sandblaster would be refilled. This allowed the sandblaster to be nearly full during all parts of the roughening process, keeping the mass flow rate from the sandblaster relatively constant. We used 50 $\mu$m aluminum oxide sand grains, and the sandblaster pressure was set to 60 psi.  To produce multiple different roughening conditions on the same sample, we used Kapton tape to block the sand grains from striking certain regions. After sandblasting, the samples were sprayed with compressed air to remove any residual sand grains from the surface. All roughening was performed within a fumehood. 
\clearpage

\section{AFM images and roughness metrics}
\label{sec:a4}
\begin{minipage}{\textwidth}
    \centering
    \includegraphics[width=1\textwidth]{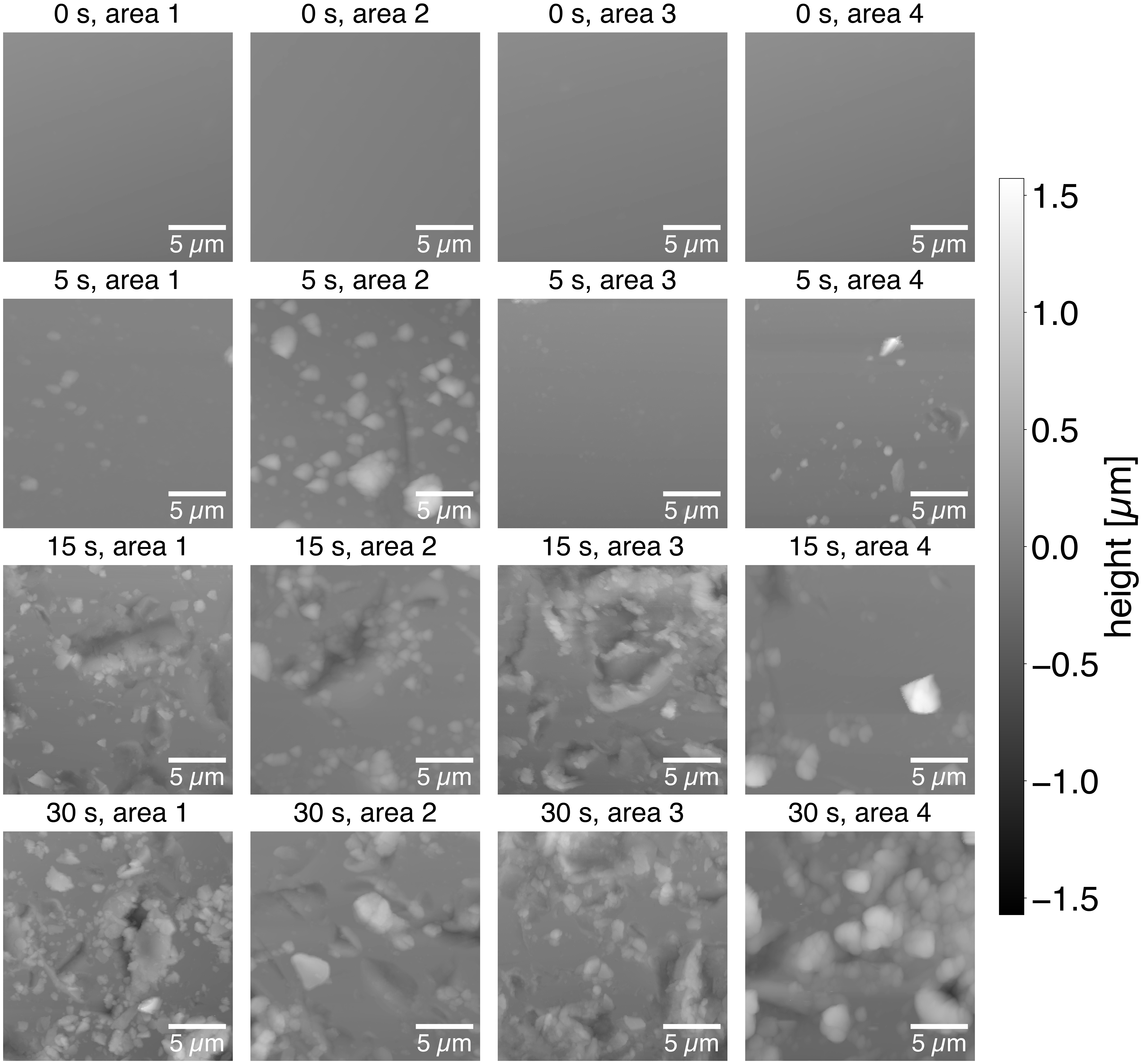}

    \captionof{figure}{
    \textbf{AFM height maps.}
    Sample height maps obtained via AFM. For each condition the height maps from four AFM scans were aggregated to determine the microfacet characteristics of that condition.
    }
    \label{fig:afm_images}
\end{minipage}

\begin{table}[tbp]
    \centering
    \setlength{\tabcolsep}{7pt}
    \begin{tabular}{@{}r *{5}{S[table-format=3.2]}@{}}
        \toprule
        & \multicolumn{5}{c}{RMS roughness [nm]} \\
        \cmidrule(l){2-6}
        Roughening time [s]
        & {Area 1} & {Area 2} & {Area 3} & {Area 4} & {Pooled} \\
        \midrule
         0 &  87.61 &  34.44 &  71.16 &  87.61 &  73.48 \\
         5 &  45.90 & 184.98 &  99.02 & 111.03 & 120.89 \\
        15 & 169.05 & 124.44 & 230.04 & 193.58 & 183.34 \\
        30 & 224.29 & 176.20 & 201.50 & 262.51 & 218.44 \\
        \bottomrule
    \end{tabular}
    \caption{AFM surface roughness for each roughening condition.
    RMS values were calculated from corrected, individually
    mean-centered height maps. Pooled values combine all pixels
    from the four scans within each condition. Each of the scans covers a 20 $\mu$m $\times$ 20 $\mu$m area and is 256 $\times$ 256 pixels.}
    \label{tab:afm_rms_roughness}
\end{table}

\clearpage
\section{Monte Carlo simulation details}
In this section we provide details on the Monte Carlo simulations used throughout this work.

Every simulation is defined by a roughness profile of the scintillator and an absorption mask that determines where in the scintillator X-rays are incident. Therefore, to simulate a given experiment, we need to encode the roughness profile and the absorption mask. The roughness profile is measured using atomic force microscopy (AFM). This provides a nanoscale map of the height variation over the scan area. By collecting multiple scans across the sample, we can determine the distribution of heights for the roughening condition being analyzed. The absorption mask informs the region of the scintillator where photons can originate. For instance, when simulating outcoupling enhancement due to surface roughness in the main text, no absorption mask is used because scintillation photons can be generated throughout the entire sample. In contrast, when simulating the blurring of an edge, photons are restricted to only originating in one half of the scintillator to mimic the edge blocking incident X-rays.

Given the sample roughness distribution and the absorption mask, we can run a Monte Carlo simulation to determine the image of the scintillator formed on the detector. We modeled optical-photon transport through the scintillator using a stochastic geometric-optics ray-tracing framework. The scintillator is represented as a homogeneous rectangular volume

\begin{equation}
\Omega=[0,W]\times[0,L]\times[0,H],
\end{equation}

where \(\Omega\) is the scintillator volume, and \(W\), \(L\), and \(H\) are its width, length, and thickness along the Cartesian \(x\), \(y\), and \(z\) axes, respectively. All coordinates and dimensions are expressed in the same length units. The scintillator has refractive index \(n_{\mathrm{s}}\) and is surrounded by a medium of refractive index \(n_{\mathrm{o}}=1\) (air). Each ray is characterized by its position \(\mathbf{r}\), whose components are the coordinates \(x\), \(y\), and \(z\), and its dimensionless unit propagation direction \(\boldsymbol{\omega}\). 
\subsection{Photon generation}
Optical emission directions are sampled isotropically:

\begin{equation}
p(\boldsymbol{\omega})=\frac{1}{4\pi},
\end{equation}

where \(p(\boldsymbol{\omega})\) is the emission-direction probability density per unit solid angle. This distribution is implemented by sampling

\begin{equation}
\omega_z\sim\mathcal{U}(-1,1),
\qquad
\phi\sim\mathcal{U}(0,2\pi),
\end{equation}

where \(\omega_z\) is the dimensionless \(z\) component of the unit propagation direction, \(\phi\) is the emission azimuth about the \(z\) axis measured from the positive \(x\) axis, and \(\mathcal{U}(x_1,x_2)\) denotes a continuous uniform distribution with lower bound \(x_1\) and upper bound \(x_2\). The propagation direction is then constructed as

\begin{equation}
\boldsymbol{\omega}
=
\left(
\sqrt{1-\omega_z^2}\cos\phi,\,
\sqrt{1-\omega_z^2}\sin\phi,\,
\omega_z
\right).
\end{equation}

The lateral distribution of emission sites depends on the absorption mask. For uniform irradiation, \(x\sim\mathcal{U}(0,W)\) and \(y\sim\mathcal{U}(0,L)\). The creation depth is sampled from the X-ray absorption profile, conditioned on absorption within the scintillator:

\begin{equation}
p(z)
=
\frac{\exp(-z/\Lambda_{\mathrm{X}})}
{\Lambda_{\mathrm{X}}\left[1-\exp(-H/\Lambda_{\mathrm{X}})\right]},
\qquad 0\le z\le H,
\end{equation}

where \(p(z)\) is the conditional creation-depth probability density, with units of inverse length, and \(\Lambda_{\mathrm{X}}\) is the X-ray attenuation length. Depth \(z\) is measured from the X-ray incident surface at \(z=0\). For the plastic scintillators considered in this work, the X-ray attenuation length is much larger than the scintillator thickness:

\begin{equation}
\epsilon
=
\frac{H}{\Lambda_{\mathrm{X}}}
\ll 1,
\end{equation}

where \(\epsilon\) is the dimensionless thickness-to-attenuation-length ratio. Expanding the exponential in this limit gives

\begin{equation}
1-\exp\left(-\frac{H}{\Lambda_{\mathrm{X}}}\right)
=
\frac{H}{\Lambda_{\mathrm{X}}}
-
\frac{H^2}{2\Lambda_{\mathrm{X}}^2}
+
\mathcal{O}\left(\epsilon^3\right).
\end{equation}

Consequently, the normalized absorption-depth distribution becomes

\begin{equation}
p(z)
=
\frac{1}{H}
\left[
1+\frac{H/2-z}{\Lambda_{\mathrm{X}}}
+
\mathcal{O}\left(\epsilon^2\right)
\right],
\end{equation}

and therefore approaches a uniform distribution:

\begin{equation}
\lim_{\Lambda_{\mathrm{X}}/H\rightarrow\infty}p(z)
=
\frac{1}{H},
\qquad 0\le z\le H.
\end{equation}

Thus, when \(\Lambda_{\mathrm{X}}\gg H\), photon creation is effectively uniform through the sample thickness; the first-order correction produces a small excess of events near the incident surface. Note that our model samples optical emission sites but does not model scintillation yield and energy deposition.

\subsection{Bulk propagation and boundary interactions}
Between boundary interactions, rays are propagated along straight trajectories,

\begin{equation}
\mathbf{r}(s)=\mathbf{r}_0+s\boldsymbol{\omega},
\qquad s>0,
\end{equation}

where \(\mathbf{r}_0\) is the starting position and \(s\) is the distance traveled along the segment. The propagation direction remains constant between boundary interactions.

The next interaction point is the intersection with the nearest of the six axis-aligned scintillator boundaries (the faces of a rectangular prism). Optical attenuation in the bulk is modeled by Beer--Lambert survival. For a segment of length \(\ell\),

\begin{equation}
P_{\mathrm{surv}}(\ell)
=
\exp\left(-\frac{\ell}{\Lambda_{\mathrm{opt}}}\right),
\end{equation}

where \(P_{\mathrm{surv}}\) is the survival probability and \(\Lambda_{\mathrm{opt}}\) is the optical attenuation length. For each propagation between boundaries, a sample from a uniform random distribution on the interval from zero to one is generated. If this value exceeds \(P_{\mathrm{surv}}\) then the ray is terminated. No bulk scattering or wavelength conversion is included.

For incidence from inside the scintillator, let \(\mathbf{n}\) be the outward unit normal pointing into the surrounding medium and \(\boldsymbol{\omega}_i\) the incident unit propagation direction. The incidence cosine is

\begin{equation}
c_i=\boldsymbol{\omega}_i\cdot\mathbf{n}.
\end{equation}

Snell's law is evaluated using

\begin{equation}
\sin^2\theta_t
=
\left(\frac{n_{\mathrm{s}}}{n_{\mathrm{o}}}\right)^2
\left(1-c_i^2\right),
\end{equation}

where \(\theta_t\) is the transmitted angle measured from the outward normal in the surrounding medium. If \(\sin^2\theta_t>1\), the photon is totally internally reflected. Otherwise, the unpolarized Fresnel reflectance is

\begin{equation}
R
=
\frac{R_s+R_p}{2},
\end{equation}

where \(R\) is the dimensionless unpolarized power reflectance, and \(R_s\) and \(R_p\) are the dimensionless power reflectances for electric fields perpendicular and parallel to the plane of incidence, respectively. The plane of incidence is spanned by the incident ray direction and the interface normal. The polarized reflectances are

\begin{equation}
R_s
=
\left(
\frac{n_{\mathrm{s}}c_i-n_{\mathrm{o}}c_t}
     {n_{\mathrm{s}}c_i+n_{\mathrm{o}}c_t}
\right)^2,
\qquad
R_p
=
\left(
\frac{n_{\mathrm{o}}c_i-n_{\mathrm{s}}c_t}
     {n_{\mathrm{o}}c_i+n_{\mathrm{s}}c_t}
\right)^2,
\end{equation}

where \(c_t\) is the cosine of the transmitted angle, given by

\begin{equation}
c_t
=
\sqrt{
1-
\left(\frac{n_{\mathrm{s}}}{n_{\mathrm{o}}}\right)^2
(1-c_i^2)
}.
\end{equation}

This real-valued expression applies when total internal reflection does not occur.

Reflection or transmission is selected stochastically with probabilities \(R\) and \(1-R\), respectively. The reflected direction is

\begin{equation}
\boldsymbol{\omega}_r
=
\boldsymbol{\omega}_i
-
2(\boldsymbol{\omega}_i\cdot\mathbf{n})\mathbf{n},
\end{equation}

where \(\boldsymbol{\omega}_r\) is the reflected unit propagation direction. The transmitted direction is

\begin{equation}
\boldsymbol{\omega}_t
=
\eta\boldsymbol{\omega}_i
-
\left(\eta c_i-c_t\right)\mathbf{n},
\qquad
\eta=\frac{n_{\mathrm{s}}}{n_{\mathrm{o}}},
\end{equation}

where \(\boldsymbol{\omega}_t\) is the transmitted unit propagation direction and \(\eta\) is the incident-to-transmitted refractive-index ratio. For incidence from the surrounding medium, the incident and transmitted refractive indices are interchanged and the interface normal is reversed.

\subsection{Microfacet surface roughness model}

We model the rough surface as an ensemble of locally smooth microfacets with different orientations, following microfacet theory \cite{walter2007microfacet}. The microfacet orientations are determined directly from the AFM measurements for each roughening condition. The underlying projected-area distribution of local unit normals is
\begin{equation}
p(\mathbf{m}) = D(\mathbf{m})(\mathbf{n}\cdot\mathbf{m}),
\qquad \mathbf{n}\cdot\mathbf{m}>0,
\end{equation}
where $\mathbf{m}$ is a dimensionless local outward unit microfacet normal, $p(\mathbf{m})$ is the underlying projected-area probability density of these normals per unit solid angle, $D(\mathbf{m})$ is the microfacet normal-distribution function per unit solid angle, and $\mathbf{n}=(0,0,1)$ is now the macroscopic outward normal of the rough surface. The condition $\mathbf{n}\cdot\mathbf{m}>0$ restricts the underlying normals to the outward-facing hemisphere. The factor $\mathbf{n}\cdot\mathbf{m}$ accounts for the projected area of each microfacet.

To construct the measured distributions, we first subtract a best-fit quadratic surface from each AFM height map to remove overall tilt and curvature. We then calculate local surface slopes using finite differences with the physical AFM pixel spacing. The dimensionless slope magnitude and corresponding microfacet tilt angle are
\begin{equation}
s = \sqrt{
\left(\frac{\partial z}{\partial x}\right)^2
+
\left(\frac{\partial z}{\partial y}\right)^2
},
\qquad
\theta_m = \arctan s,
\end{equation}
where $z(x,y)$ is the detrended AFM surface height, $x$ and $y$ are lateral AFM coordinates expressed in the same length units as $z$. $\partial z/\partial x$ and $\partial z/\partial y$ are the dimensionless local slopes along these coordinates, $s$ is their dimensionless resultant magnitude, and $\theta_m$ is the microfacet tilt angle between the local and macroscopic outward normals. In this subsection, $z(x,y)$ denotes surface height rather than photon creation depth, and $s$ denotes slope magnitude rather than propagation distance.

Slope magnitudes from the AFM scans for each roughening condition are pooled, with each pixel contributing equally. The resulting histogram represents the underlying slope distribution.

We assume isotropic surface roughness, so the underlying slope azimuth is uniformly distributed. A local normal drawn from this underlying distribution is therefore
\begin{equation}
\mathbf{m}
=
\frac{(-s\cos\phi,\,-s\sin\phi,\,1)}
{\sqrt{1+s^2}},
\qquad
\phi\sim\mathcal{U}(0,2\pi),
\end{equation}
where $\phi$ now denotes the azimuth of the surface-height gradient measured from the positive $x$ axis. Equations S19 and S21 describe the underlying surface statistics. During the microsurface random walk, encountered normals are sampled from the corresponding direction-dependent visible-normal distribution, which weights facets according to their projected area along the incident ray direction.

Multiple scattering within the rough interface is modeled using a dielectric Smith random walk following Heitz et al \cite{heitz2016multiple}. At each local encounter, a microfacet normal is sampled from the direction-dependent visible-normal distribution derived from the measured AFM slope histogram, and reflection or transmission is selected using the Fresnel probabilities. Rays that encounter additional microfacets continue scattering until they escape the microsurface into either the scintillator or the surrounding medium. The encounter and escape probabilities are determined by the Smith shadowing function calculated from the same AFM histogram.

Reflection and refraction at each local encounter are calculated using the Fresnel relations in Section S3.B, with the sampled microfacet normal replacing the macroscopic normal and the refractive indices assigned according to the medium occupied by the ray. Masking and shadowing are incorporated through the Smith random walk’s encounter and escape probabilities. 

For a unit propagation direction $\boldsymbol{\omega}$, we define
the absolute direction cosine
\begin{equation}
\mu = |\boldsymbol{\omega}\cdot\mathbf{n}|,
\end{equation}
where $\mu$ ranges from zero for grazing propagation to unity for normal propagation. The single-direction Smith visibility factor is
\begin{equation}
G_1(\boldsymbol{\omega},\mathbf{m})
=
\begin{cases}
\displaystyle \frac{1}{1+\Lambda(\mu)},
&
(\boldsymbol{\omega}\cdot\mathbf{m})
(\boldsymbol{\omega}\cdot\mathbf{n})>0,
\\[6pt]
0, & \text{otherwise},
\end{cases}
\end{equation}
where $G_1$ is the dimensionless visibility factor and $\Lambda(\mu)$ is the dimensionless Smith shadowing function. The sign condition requires agreement
between the sides of the local and macroscopic interfaces
associated with the propagation direction.

For our isotropic empirical slope distribution, the Smith
shadowing function is evaluated directly from the AFM histogram:
\begin{equation}
\Lambda(\mu)
=
\sum_{j=1}^{B}
h_j\,\mathcal{L}\left(
s_j\frac{\sqrt{1-\mu^2}}{\mu}
\right),
\qquad 0<\mu\leq 1,
\end{equation}
where $B$ is the number of slope bins, $j$ indexes those bins, $h_j$ is the fraction of pooled AFM pixels in bin $j$, $s_j$ is its dimensionless slope-magnitude center, and $\mathcal{L}$ is the dimensionless azimuth-averaged excess projected-area function defined below.

For uniform azimuth, the dimensionless function $\mathcal{L}$
has the analytical form
\begin{equation}
\mathcal{L}(a)
=
\begin{cases}
0, & 0\leq a\leq 1,
\\[4pt]
\displaystyle
\frac{\sqrt{a^2-1}-\arccos(1/a)}{\pi},
& a>1,
\end{cases}
\end{equation}
where $a$ is the nonnegative dimensionless argument appearing in the histogram sum. This expression averages the positive
part of the projected slope in excess of unity over all
azimuthal orientations.

Thus, both the sampled microfacet orientations and their Smith
visibility factors are determined by the same measured slope
histogram. The visibility factors are tabulated before ray
tracing and interpolated during propagation.

\subsection{Collection and image formation}
After a ray exits the scintillator, we calculate its contribution to the final image. Since the experiment uses an imaging objective with a finite numerical aperture, only rays that exit within the acceptance angle contribute to the image. For collection into air with the optical axis along the positive $z$ direction, this condition is

\begin{equation}
\omega_{t,z}
\ge
\cos\theta_{\max}
=
\sqrt{1-\mathrm{NA}^2}.
\end{equation}

Here, \(\omega_{t,z}\) is the \(z\) component of \(\boldsymbol{\omega}_t\) after the ray leaves the microsurface, \(\theta_{\max}\) is the objective's maximum collection half-angle measured from the positive \(z\) axis, and \(\mathrm{NA}\) is the objective's numerical aperture.

If a ray satisfies this condition then it is included in the detector image at the location where it exited the scintillator. Performing many Monte Carlo trials allows us to build up an image on the detector. This image is then used to calculate downstream metrics such as the outcoupling enhancement or remote outcoupling strength. 

One additional step is included for the simulation of the straight edge being imaged through the scintillator. This case is sensitive to blurring throughout the imaging system. To capture this we measured the intrinsic blurring due to the imaging optics themselves. For our imaging optics we estimated this to be roughly 70 $\mu$m of Gaussian blur, so this was added to the simulated detector image for the imaging resolution simulation.

\clearpage

\section{Scintillator imaging setup}

\begin{minipage}{\textwidth}
    \centering
    \includegraphics[width=0.93\textwidth]{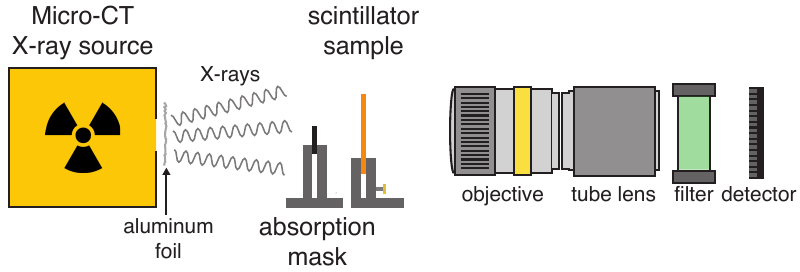}

    \captionof{figure}{
    \textbf{Scintillator imaging setup.}
    X-rays propagate from the micro-CT source through the aluminum foil. They then are selectively absorbed by the absorption mask. The shape of the mask depends on the specific experiment. The transmitted X-rays then strike the sample where they generate visible photons. These photons are collected by an imaging objective connected to a tube lens. A filter is used before the detector to isolate the scintillation signal. 
    }
    \label{fig:scintillator_imaging}
\end{minipage}
\vspace{1mm}

The same X-ray imaging setup was used for each of the X-ray experiments. These are the measurements of the outcoupling enhancement, imaging resolution, and remote outcoupling. 

The X-rays are generated by the X-ray source from the ZEISS Xradia 620 Versa micro-CT machine. To remove low energy photons from the beam, aluminum foil was placed over the source. The X-rays travel to the absorption mask. The geometry of the mask depends on the experiment, but in each case the appropriate part of the beam was blocked by a 5.6 mm thick stainless steel block. Black tape was placed over the block to prevent reflections off the steel. The scintillator sample was mounted downstream of the absorption mask. To collect images of the scintillator, the detector stage was focused on the back surface of the sample. We used a Mitutoyo Plan Apo Infinity Corrected 5x objective. We attached a filter after the objective to isolate the scintillation signal. For BC-430 we used a FF01-565/133-25 filter, while for BC-428 we used FF01-550/200-25. Detection was performed using the detector built into the micro-CT system. In the table below are the X-ray tube parameters used for each of the X-ray experiments reported in this work. 

\begin{table}[tbp]
    \centering
    \setlength{\tabcolsep}{7pt}
    \begin{tabular}{@{}l
        S[table-format=3.0]
        S[table-format=3.1]
        S[table-format=3.0]
        S[table-format=1.0]@{}}
        \toprule
        & \multicolumn{4}{c}{X-ray acquisition settings} \\
        \cmidrule(l){2-5}
        Experiment & {Tube voltage [kV]} & {Tube current [$\mu$A]}
        & {Exposure [s]} & {Binning factor} \\
        \midrule
        Intensity          & 120 & 145.8 &  30 & 2 \\
        Resolution         &  50 &  90.0 &  60 & 1 \\
        USB stick          & 120 & 145.8 & 240 & 2 \\
        Fruit fly          &  30 &  66.7 &  90 & 4 \\
        Remote outcoupling &  30 &  66.7 & 240 & 4 \\
        \bottomrule
    \end{tabular}
    \caption{X-ray experiment acquisition settings. Exposure times are given per image.}
    \label{tab:xray_acquisition_settings}
\end{table}

\clearpage

\section{Sample consistency}
\begin{minipage}{\textwidth}
    \centering
    \includegraphics[width=1\textwidth]{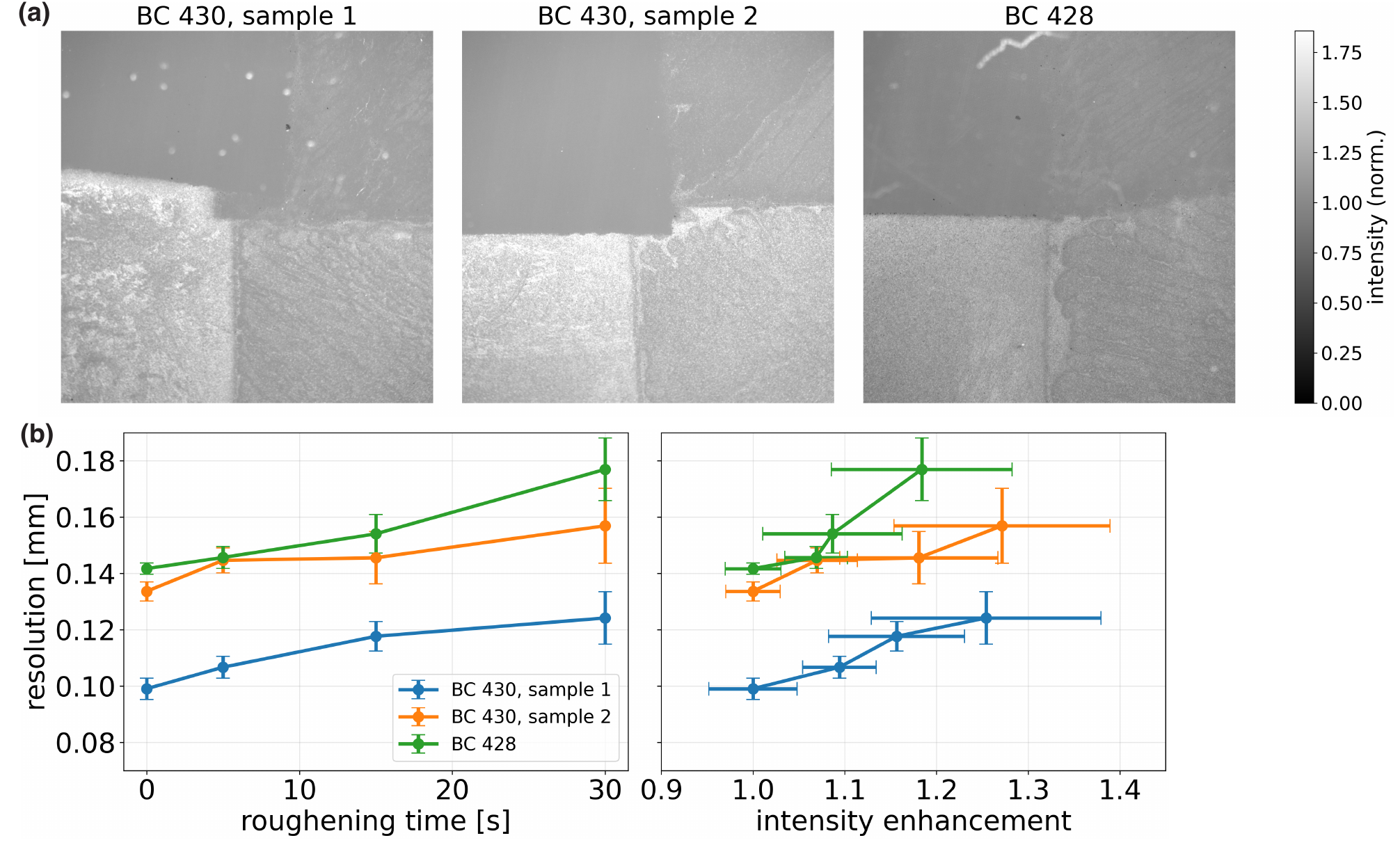}

    \captionof{figure}{
    \textbf{Sample consistency.}
    \textbf{(a)} Experimental X-ray images of scintillator surface for three samples. BC-430 sample 1 is the plastic scintillator whose data is reported in the main text. BC-428 is a similar plastic scintillator whose emission wavelength is blue-shifted relative to BC-430. 
    \textbf{(b)} Imaging resolution as a function of roughening time and intensity enhancement for the three different samples. 
    }
    \label{fig:sample_consistency}
\end{minipage}
\vspace{1mm}

In order to determine the repeatability of our results we prepared multiple samples using the same roughening technique and performed the same measurements on them. Figure \ref{fig:sample_consistency} shows X-ray data from three samples: two BC-430 and one BC-428. BC-428 is also a plastic scintillator and mainly differs from BC-430 by its emission wavelength. We see that the X-ray data is largely consistent across the samples, indicating that the roughening procedure is repeatable. All three samples display the same trends in imaging resolution as a function of the roughening time and intensity enhancement. Note that differences in the samples present before roughening could account for some of the disparity between the samples.  

\clearpage

\bibliography{main}